\documentclass[twocolumn]{article}

\usepackage{graphics}
\usepackage{graphicx}
\usepackage{amssymb}
\usepackage{natbib}

\title{Collisionless Magnetic Field Reconnection From First Principles: What It Can and Cannot Do}
\author{F. S. Mozer\\
Physics Department and Space Sciences Laboratory,\\
University of California, Berkeley, CA 94720\\
fmozer@ssl.berkeley.edu}
\date{\quad}

\begin{document}

\maketitle
{\small
\parskip=12pt
\parindent=0pt

{\bf Abstract.} The underlying concepts of collisionless magnetic field reconnection in the vicinity of the diffusion regions are developed qualitatively from first principles.  One conclusion of this analysis is that energetic ions and electrons observed in the terrestrial magnetosphere or from the sun are not accelerated to the observed energies in the diffusion regions.\\
}

 In plasmas from the laboratory to the terrestrial magnetosphere, the sun, and accretion disks in astrophysics, it is observed that magnetic field geometries change and particles are accelerated on time scales that are short compared to magnetic field diffusion times due to classical Coulomb collisions.  These observations require physical processes different from Coulomb resistivity.  One such process is collisionless magnetic field reconnection, which is discussed in this paper by defining boundary conditions in a region of space and applying first principles inside these boundaries to qualitatively deduce the plasma flow, energy flow and the fields associated with reconnection.  The goals of this approach are to remove some of the mystery and confusion that may surround field line motion and magnetic field reconnection. 

     Magnetic field reconnection is defined as the process that occurs when magnetized plasmas with different magnetic field orientations flow together to alter the connectivity of the magnetic field lines.  It occurs on spatial scales of the electron and ion skin depths ($c/\omega_{pe}$ and $c/\omega_{pi}$, where $\omega_{pe}$ and $\omega_{pi}$ are the ion and electron plasma frequencies, respectively), \citep{vasyliunas75} after which accelerated plasma expands into a larger region where it may be further modified or accelerated by processes that are beyond the scope of the present discussion. 

     We begin with the physical picture of Fig.\ 1, which depicts a region a few ion skin depths ($\sim$100 km at the sub-solar magnetopause) in size.  If the figure describes the sub-solar region, the vertical direction in the figure is perpendicular to the plane of the ecliptic and the left side of the figure is in the sunward direction.  The plasma that fills the figure is a two component neutral fluid that is bounded by magnetic field lines that are up on the right side and down on the left side of the figure, and for which, in each case, there is an electric field, $E_{o}$, out of the plane of the figure.  Thus, the plasma on both sides of the figure $\mathbf{E}_{o} \times \mathbf{B} / B^{2}$ drifts towards the center of the figure.  The two oppositely directed magnetic fields may be unequal in magnitude or a magnetic field component, called the guide field, may exist out of the plane of Fig.\ 1.  These features add complexity to the geometry without significantly modifying the basic physical processes that occur inside a few ion skin depths, so the guide field is assumed to be zero and the oppositely directed magnetic fields are assumed to be equal in magnitude.  Fig.\ 1 describes the boundary conditions for the processes that occur in the central region of the figure and that will be deduced in the following discussion.

\begin{figure}
\centering \includegraphics[width=2.3in]{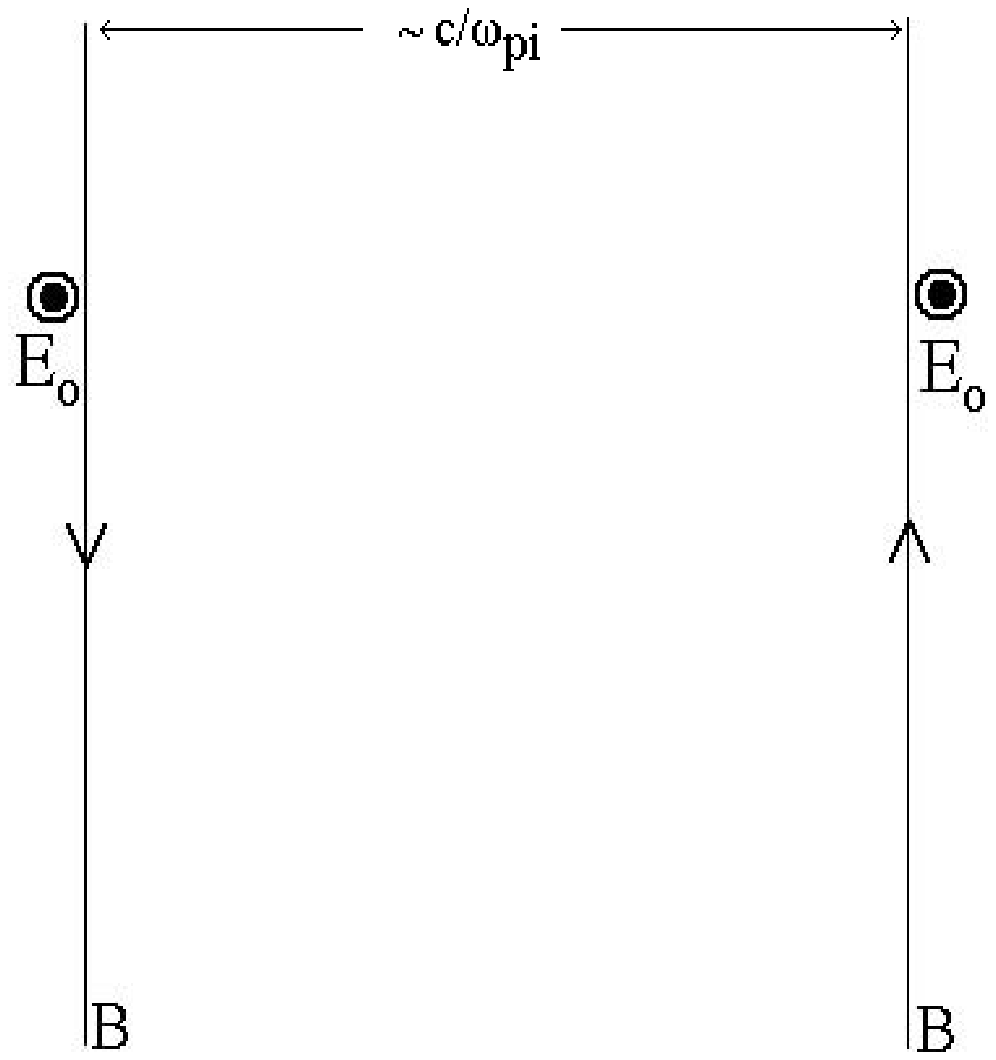}\\
\centering Figure 1
\label{fig01}
\end{figure}

     The physics of reconnection depends on the electric field component out of the plane of Fig.\ 1 at the center of the figure, which is sometimes called the tangential electric field.  If it is zero, the two plasmas flow around each other into or out of the plane of the figure because there is no $\mathbf{E}_{o} \times \mathbf{B} / B^{2}$ flow in the plane of the figure in this central region.  This would be the case for a closed terrestrial magnetosphere in which the solar wind plasma flows around the obstacle much as water flows around a rock in a stream.  On the other hand, if the tangential electric field is non-zero, the plasmas continue flowing towards each other into the central region of the figure and magnetic field reconnection occurs as discussed below.  The main effort in theoretical reconnection research has been the search for a large value of the tangential electric field in the central region \citep{parker57,parker63,sonnerup84,birn01,shay01}.  Because this research is beyond the scope of this article, it is assumed that the tangential electric field in the central region is equal to $E_{o}$.\\

     \textbf{Why is there so much theoretical interest in making $\mathbf{E_{o}}$ large?}  Consider the current out of the plane in the central region of Fig.\ 1, which is required by the curl of the magnetic field.  A non-zero electric field results in a positive value of $\mathbf{j} \cdot \mathbf{E}$ in the central region of the figure and conversion of electromagnetic energy into particle energy, which is what magnetic field reconnection is all about.\\

     \textbf{Where does the energy associated with this positive $\mathbf{j} \cdot \mathbf{E}$ originate?}  To answer this question, an analogy will be made with the electric circuit of Fig.\ 2, consisting of a battery and a resistor.  From freshman physics, the electromagnetic energy conversion rate is VI where V is the battery voltage and I is the current.  A second method for computing the energy conversion is to integrate $\mathbf{j} \cdot \mathbf{E}$ over the resistor volume. With $\mathbf{j}$ integrated over $\pi r^{2}$, the cross-sectional area of the resistor, to give I and E integrated over the length of the resistor to become the potential drop V, the energy conversion rate is again VI.  A third method is to consider the integral of the Poynting flux through the surface that surrounds the resistor.  Because $E$ is everywhere parallel to the axis of the cylinder and $B$ everywhere circles the resistor, the Poynting flux, $\mathbf{E} \times \mathbf{B} / \mu_{o}$, is everywhere inward and its integrated magnitude is $E h (2 \pi r B ) / \mu_{o}$).  Because $Eh = V$ and $2 \pi r B / \mu_{o} = I$, the result is again VI.  A fourth way to compute the power dissipation is to utilize a later proof in this paper to say that magnetic field lines carry energy density $B^{2} / \mu_{o}$ into the resistor at the $\mathbf{E}_{o} \times \mathbf{B} / B^{2}$ velocity, where the magnetic field is annihilated.  This field line motion inserts $(B^{2} / \mu_{o})(E/B)$ or $EB / \mu_{o}$ units of energy per unit area per second, which is the same power input as given by the Poynting flux, so the answer is again VI.

\begin{figure}
\centering \includegraphics[width=2.3in]{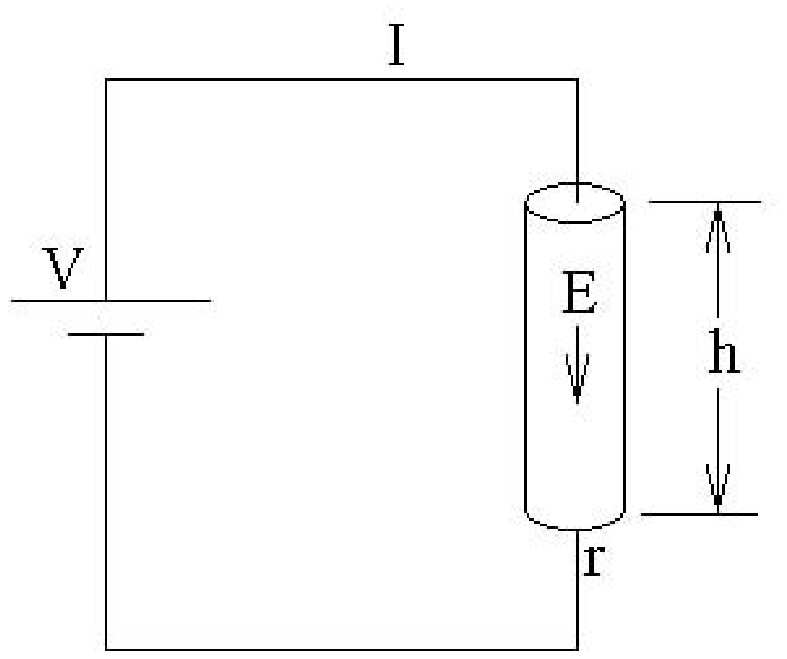}\\
\centering Figure 2
\label{fig02}
\end{figure}

     If the EMF (in the case of Fig.\ 2, a battery) is constant, the energy conversion rate is constant.  This is analogous to the case of steady state reconnection at the dayside magnetopause, where the energy is continuously supplied by the solar wind as it slows down.  In both cases, the partial derivatives of all fields with respect to time are zero, so it is confusing to think of this process as magnetic field annihilation.  This is just one example of problems arising from misinterpretation of the concept of moving magnetic field lines.\\

      \textbf{Where does the converted electromagnetic energy go?}  It may generate waves or heat and/or accelerate the plasma.  Assuming that all the available energy goes into particle acceleration, the particle energization is computed in a static model of a single ion species as follows.  Consider a closed surface that extends between the two magnetic field lines of Fig.\ 1 and that has a length, $h$, along the field lines and width, $w$, out of the figure.  There is a net Poynting flux, $2whE_{o}B/\mu_{o}$, through this surface from the right and the left.  The number of particles that move into this volume per second at the $\mathbf{E}_{o} \times \mathbf{B} / B^{2}$ velocity is $2whnE_{o}/B$, where $n$ is the plasma density.  Thus, the energy gained per particle is $(2whE_{o}B / \mu_{o}) / (2whnE_{o} / B) = B^{2} / n\mu_{o}$, and the average velocity increase of a low energy ion is $v = [2B^{2}/mn\mu_{o}]^{1/2}$, where $m$ is the ion mass.  Thus, within a factor $\sim$1, the bulk flow of low energy ions increases at most to the Alfven speed in the diffusion region after reconnection.  This has many consequences:

\begin{itemize}
\item The flux, \emph{not the energy}, of the ions emerging from the diffusion region increases with the increasing magnitude of the reconnection electric field, $E_{o}$.  The ion energy is limited to that associated with the Alfven speed which, for typical parameters at the sub-solar reconnection region is $\sim$1 keV.
\item The electron bulk speed is also increased at most to the ion Alfven speed to maintain neutrality in the absence of an outgoing current.  However, the electron energy is about 1/1856 of the ion energy because of the mass ratio so electrons are much less accelerated than ions in the diffusion region.
\item Although the energies of particles leaving the diffusion region are less than those observed in space, such particles may be accelerated further after leaving the diffusion regions by utilizing the particle energy resulting from reconnection \citep{birnhesse05} or by using additional energy sources and mechanisms, such as are available by surfing on plasma waves, particles following meandering orbits, etc.
\end{itemize}

     \textbf{What happens to the plasma that is accelerated by the electromagnetic energy conversion?}  It must leave the central region by either flowing up/down or into/out of the plane of Fig.\ 1.  A convenient simplification results from assuming that reconnection is a two-dimensional process, from which it follows that the plasma exits the central region of Fig.\ 1 in the vertical direction.  It is not necessary that the geometry be two-dimensional and in the real world it isn't \citep{mozer03b}, although the current, \textbf{j}, flowing out of the plane of Fig.\ 1 suggests a tendency for a two-dimensional geometry.  To keep the geometry simple without losing any physics, it is assumed in this discussion that the plasma exits the central region by flowing vertically.  This leads to bulk plasma flows like those indicated by the heavy lines in Fig.\ 3.\\

\begin{figure}
\centering \includegraphics[width=2.3in]{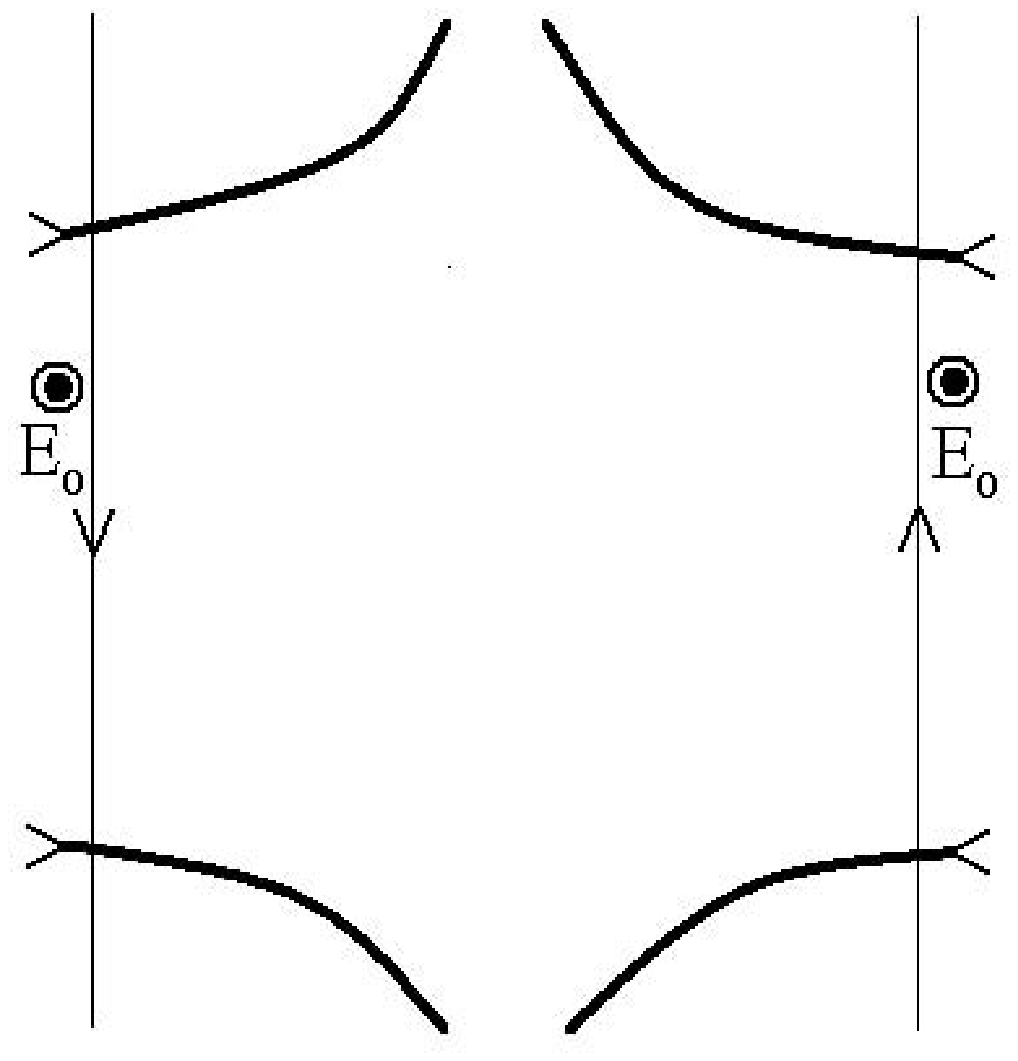}\\
\centering Figure 3
\label{fig03}
\end{figure}

     \textbf{What is the magnetic field geometry associated with the flows of Fig.\ 3?}  Because the plasma flow is perpendicular to $B$, at least initially, the magnetic field lines in the central region must curve such that they tend to be perpendicular to the flow contours of Fig.\ 3.  There are two approaches for obtaining a quantitative picture of the magnetic field geometry.  They are:

\begin{enumerate}
\item Solve Maxwell's equations.
\item Utilize a method of visualizing the magnetic field geometry without having to solve Maxwell's equations.  One such method is to assume that the magnetic field evolves in time by magnetic field lines moving at the $\mathbf{E} \times \mathbf{B} / B^{2}$ velocity.
\end{enumerate}

Method 2 is preferable to method 1 as long as method 2 gives the same solution as method 1.  If it does not, there is nothing magic about the situation. It is just that Maxwell's equations must be solved in order to determine the magnetic field.

     In the following discussion, a condition is found for which the 
magnetic field geometry obtained by assuming that field lines move with $\mathbf{E} \times \mathbf{B} / B^{2}$ is the same as that produced by Maxwell's equations \citep{longmire63}.  The concept of moving magnetic field lines implies that field lines do not just pop up in a region of increasing field strength, but that field lines flow into that region.  Because magnetic field lines are conserved in this picture, they obey a continuity equation which, in a coordinate system with the local magnetic field in the Z-direction, is 

\begin{equation}
\label{eqn01}
\partial B_{Z} / \partial t + \nabla \cdot (B \mathbf{v} ) = 0
\end{equation}

Selecting the field line velocity to be $\mathbf{v} = \mathbf{E} \times \mathbf{B} / B^{2}$, the components of $B\mathbf{v}$ are $(B\mathbf{v})_{X} = E_{Y}$ and $(B\mathbf{v})_{Y} = -E_{X}$, so equation \ref{eqn01} becomes

\begin{equation}
\label{eqn02}
\partial B_{Z} / \partial t + \partial E_{Y} / \partial x - \partial E_{X} / \partial y = 0
\end{equation}

which is the z-component of Faraday's law.  Thus, \emph{without approximation and in the presence or absence of plasma}, the magnitude of the magnetic field evolves as is required by Maxwell's equations if magnetic field lines move with the $\mathbf{E} \times \mathbf{B} / B^{2}$ velocity.  It is noted that any velocity $\mathbf{v'}$ satisfying  $\nabla \cdot (B\mathbf{v'}) = 0$ may be added to $\mathbf{E} \times \mathbf{B} / B^{2}$ without modifying equation 1.  Thus, there are an infinite number of magnetic field line velocities that preserve the magnitude of the magnetic field.    

     The direction of a field line moving with $\mathbf{E} \times \mathbf{B} / B^{2}$ must also be the same as that obtained from Maxwell's equations.  To find the condition under which this is true, consider two surfaces, S$_{1}$ and S$_{2}$ in Fig.\ 4, that are perpendicular to the field at times $t$ and $t + \delta t$.  At time $t$, a magnetic field line intersects the two surfaces at points $\mathbf{a}$ and $\mathbf{b}$.  Thus, the vector $(\mathbf{b} - \mathbf{a})$ is parallel to $\mathbf{B}(t)$.  At the later time, $t + \delta t$, point $\mathbf{a}$ has moved along S$_{1}$ at velocity $\mathbf{E} \times \mathbf{B} / B^{2}(a)$ to point $\mathbf{a'}$ and it is on the illustrated magnetic field line.  Meanwhile, point $\mathbf{b}$ has moved along S$_{2}$ to $\mathbf{b'}$ at velocity $\mathbf{E} \times \mathbf{B} / B^{2}(b)$ and it may or may not be on the field line that passes through $\mathbf{a'}$.  The question is, what is the constraint on these motions that result in $\mathbf{a'}$ and $\mathbf{b'}$ being on the same magnetic field line, i.e., that result in $(\mathbf{b'} - \mathbf{a'})$ being parallel to $\mathbf{B}(t + \delta t)$?  

\begin{figure}
\centering \includegraphics[width=2.in]{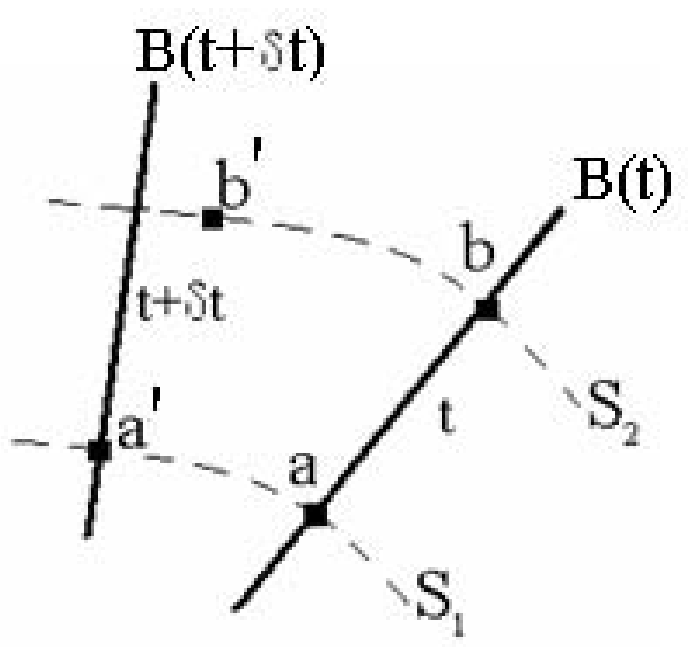}\\
\centering Figure 4
\label{fig04}
\end{figure}

     To find the constraint, we start with the identity $(\mathbf{b'} - \mathbf{a'}) = (\mathbf{b} - \mathbf{a}) + (\mathbf{b'} - \mathbf{b}) - (\mathbf{a'} - \mathbf{a})$.  The terms on the right side of this equation are:
\begin{eqnarray}
(\mathbf{b} - \mathbf{a})  & = & \varepsilon \mathbf{B} \textrm{, because } (\mathbf{b} - \mathbf{a})
					    \textrm{ is parallel to B.} \nonumber \\
(\mathbf{a'} - \mathbf{a}) & = & [\mathbf{E} \times \mathbf{B} / B^{2}(a)] \delta t \nonumber \\
(\mathbf{b'} - \mathbf{b}) & = & [\mathbf{E} \times \mathbf{B} / B^{2}(b)] \delta t \nonumber \\
          			   & = & [\mathbf{E} \times \mathbf{B} / B^{2}(a) + \nonumber \\
				   &	 &  \hspace{0.2in}(\mathbf{b} - \mathbf{a}) \,
					    \partial(\mathbf{E} \times \mathbf{B} / B^{2}(a)/ \partial r)] \delta t \nonumber \\
		               & = & [\mathbf{E} \times \mathbf{B} / B^{2}(a) + \nonumber \\
				   &   &  \hspace{0.2in} \varepsilon \mathbf{B}
					    \cdot \nabla (\mathbf{E} \times \mathbf{B} / B^{2}(a))] \delta t \nonumber 
\end{eqnarray}

\noindent where $r$ is along the magnetic field line at time $t$. Combining terms gives

\begin{equation}
\label{eqn03}
(\mathbf{b'} - \mathbf{a'}) / \varepsilon = \mathbf{B} + \mathbf{B} \cdot \nabla(\mathbf{E} \times \mathbf{B} / B^{2}) \delta t
\end{equation}

\noindent Also

\begin{eqnarray}
\mathbf{B'} & = & \mathbf{B}\left(a, t + \delta t\right) \nonumber \\
            & = & \mathbf{B} + \left(\partial \mathbf{B} / \partial t\right) \delta t + \nonumber \\
		&   & \hspace{0.2in} (\mathbf{E} \times \mathbf{B} / B^{2}) \cdot \nabla)
      		\mathbf{B} \delta t
\label{eqn04}
\end{eqnarray}

\noindent The problem reduces to finding the constraints that are imposed by the requirement that the right sides of equations \ref{eqn03} and \ref{eqn04} have a zero cross-product.  To first order in $\delta t$, this gives

\begin{eqnarray}
\mathbf{B} \times [\partial \mathbf{B} / \partial t +
(\mathbf{E} \times \mathbf{B} / B^{2}) \cdot \nabla) \mathbf{B} \, - &   & \nonumber \\
\mathbf{B} \cdot \nabla(\mathbf{E} \times \mathbf{B} / B^{2})]    & = & 0
\label{eqn05}
\end{eqnarray}

\noindent Using the vector identity for $\mathbf{\nabla} \times \mathbf{M} \times \mathbf{N}$, for any two vectors $\mathbf{M}$ and $\mathbf{N}$, allows rewriting equation \ref{eqn05} as

\begin{eqnarray}
\mathbf{B} \times
\{ \partial \mathbf{B} / \partial t +
\nabla \times ( \mathbf{B} \times (\mathbf{E} \times \mathbf{B} / B^{2})) \, +  \nonumber \\
\mathbf{B}(\nabla \cdot (\mathbf{E} \times \mathbf{B} / B^{2}) -
(\mathbf{E} \times \mathbf{B} / B^{2})\nabla \cdot \mathbf{B}\}  = 0
\label{eqn06}
\end{eqnarray}

\noindent The last term on the right is zero because $\nabla \cdot \mathbf{B} = 0$ and the next to last term may be neglected because its cross product with $\mathbf{B}$ is zero.  Because
$\mathbf{B} \times (\mathbf{E} \times \mathbf{B} / B^{2}) = \mathbf{E} - \mathbf{E}_{\parallel}$ and
$\partial \mathbf{B} / \partial t = - \nabla \times \mathbf{E}$, equation \ref{eqn06} is

\begin{equation}
\mathbf{B} \times (\nabla \times \mathbf{E}_{\parallel}) = 0
\label{eqn07}
\end{equation}

\noindent This is the constraint that makes the direction of the magnetic field during its $\mathbf{E} \times \mathbf{B} / B^{2}$ motion the same as that found from Maxwell's equations.  This condition is sometimes loosely stated as the requirement that the parallel electric field must be zero for field lines to move at the $\mathbf{E} \times \mathbf{B} / B^{2}$ velocity.
 
     The existence and/or properties of plasma did not enter into the equation \ref{eqn07} requirement that field line motion at $\mathbf{E} \times \mathbf{B} / B^{2}$ produces the same result as Maxwell's equations.  Thus, field line motion does not require the presence of plasma, contrary to what some discussions of field line motion assume.

     A vacuum example in which field line motion does not produce the same result as Maxwell's equations occurs for a magnetic dipole placed in a uniform electric field, with the dipole axis parallel to the electric field direction \citep{falthammer05}.  The $\mathbf{E} \times \mathbf{B} / B^{2}$ direction of a magnetic field line in the northern hemisphere is opposite to that on the same field line in the southern hemisphere, so the magnetic field lines appear to twist.  This result disagrees with Maxwell's equations for a dipole magnetic field, so something is wrong.  In fact, $\mathbf{B} \times (\nabla \times \mathbf{E}_{\parallel}) \neq 0$ everywhere, so the magnetic field evolution cannot be obtained by assuming that magnetic field lines move with the $\mathbf{E} \times \mathbf{B} / B^{2}$ velocity.  There is nothing mysterious about this problem.  It is just that the magnetic field must be obtained from Maxwell's equations and not from moving field lines.

     In the assumed absence of a parallel electric field in Fig.\ 3, the magnetic field evolution can be determined by moving a pair of magnetic field lines with the $\mathbf{E} \times \mathbf{B} / B^{2}$ velocity.  This produces the magnetic field geometry of Fig.\ 5, which may be thought of as a superposition of snapshots at times $t1, \cdots, t5$ of two magnetic field lines that $\mathbf{E} \times \mathbf{B} / B^{2}$ towards each other.  There is the obvious problem that the magnetic field lines in the center of the plot have passed over each other to produce points having the magnetic field in two directions.  This violation of Maxwell's equations means that there must be a parallel electric field in the central region such that the magnetic field evolution in this region cannot be obtained by any means other than solving Maxwell's equations.  There is nothing magic about the central region, but only that one cannot obtain the magnetic field geometry in this region by moving magnetic field lines.

\begin{figure}
\centering \includegraphics[width=2.3in]{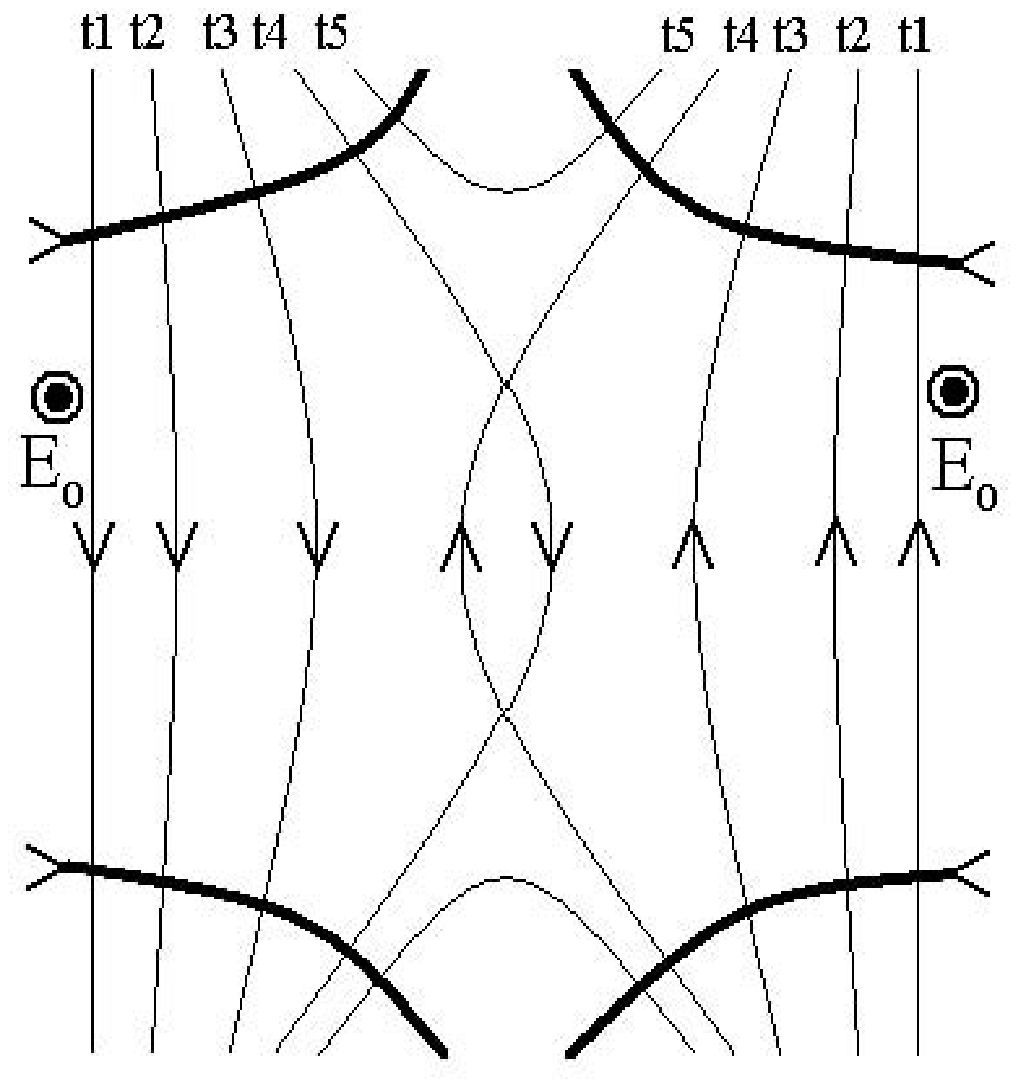}\\
\centering Figure 5
\label{fig05}
\end{figure}   

      Eliminating the magnetic field lines in the region having a parallel electric field leaves the geometry of Fig.\ 6.  The reader is reminded that  Fig.\ 6 shows the ion diffusion region while the interior of the dashed box will be shown to contain one or more electron diffusion regions.

     From the geometry of Fig.\ 6, magnetic field lines to the right of the figure in the input flow region have become connected in the outflow region to magnetic field lines that were in the left input flow region.  This is the change of magnetic field connectivity that is required by the earlier definition of magnetic field reconnection.
 
    The only use of the concept of moving magnetic field lines in the above discussion has been to aid in determining the magnetic field configuration.  If more than this use is made of moving field lines, there is a risk of drawing erroneous conclusions such as, because the magnetic field lines move in Fig.\ 6, $\partial \mathbf{B} / \partial t$ is non-zero.  This is incorrect.

\begin{figure}
\centering \includegraphics[width=2.3in]{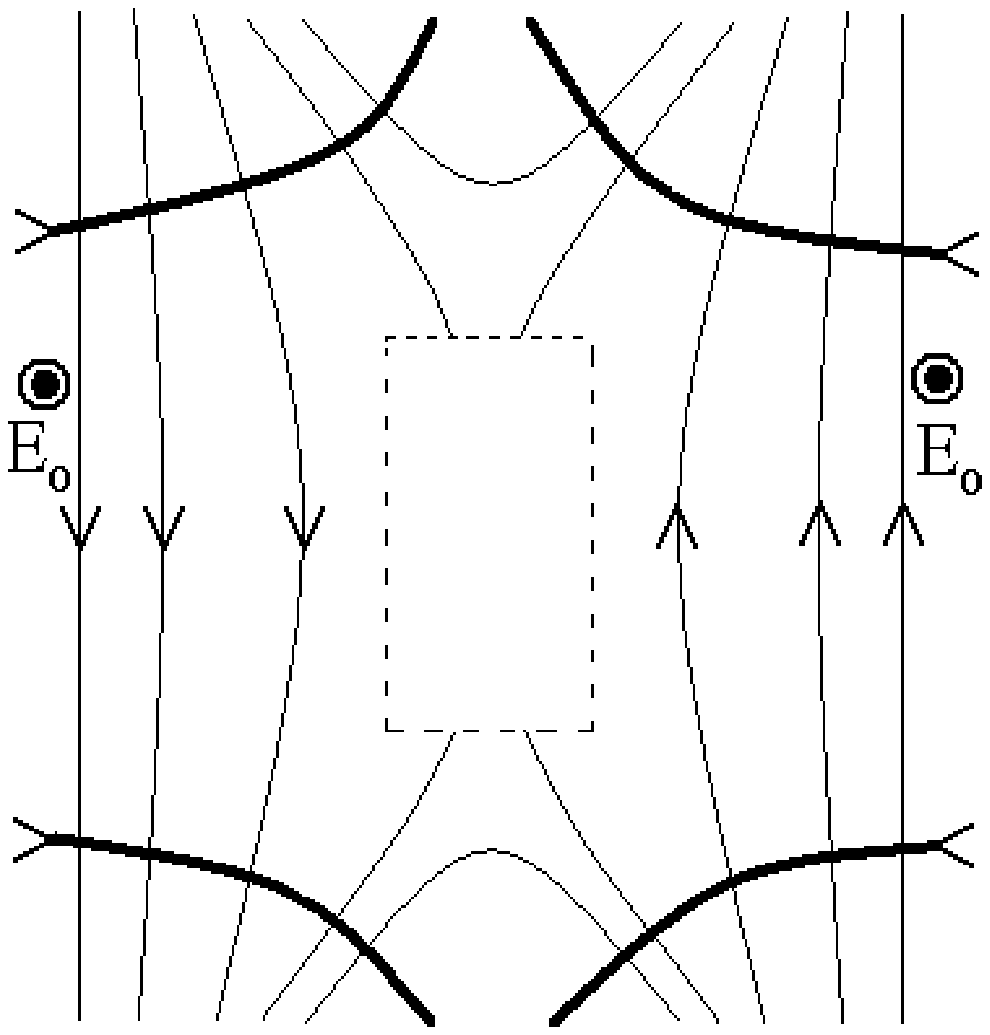}\\
\centering Figure 6
\label{fig06}
\end{figure} 

     The change of magnetic field connectivity in terrestrial reconnection is illustrated in Fig.\ 7.  This figure shows three differently connected regimes, type 1 which contains magnetic field lines that go through the sun without going through the earth, type 2 which go through the earth without going through the sun and type 3 which go through both the earth and the sun.  These topological boundaries are shown as the dashed X-line in Fig.\ 8.

\begin{figure}
\centering \includegraphics[width=2.8in]{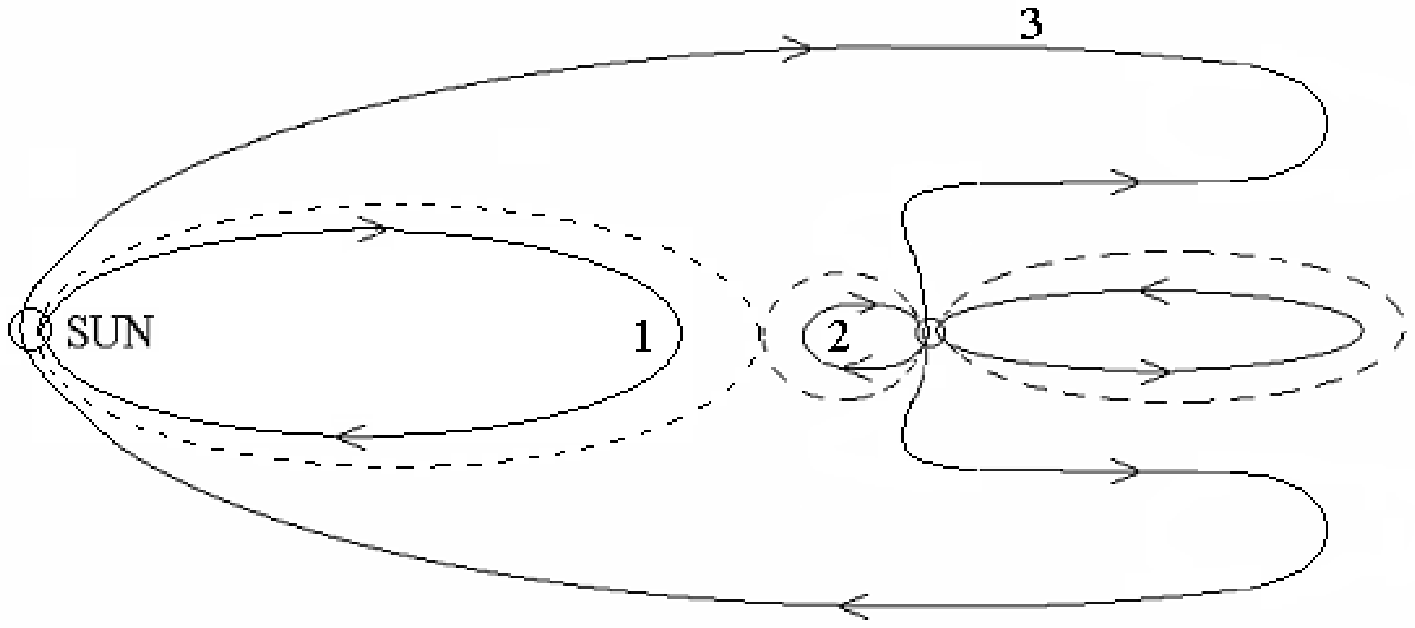}\\
\centering Figure 7
\label{fig07}
\end{figure}

\begin{figure}
\centering \includegraphics[width=2.3in]{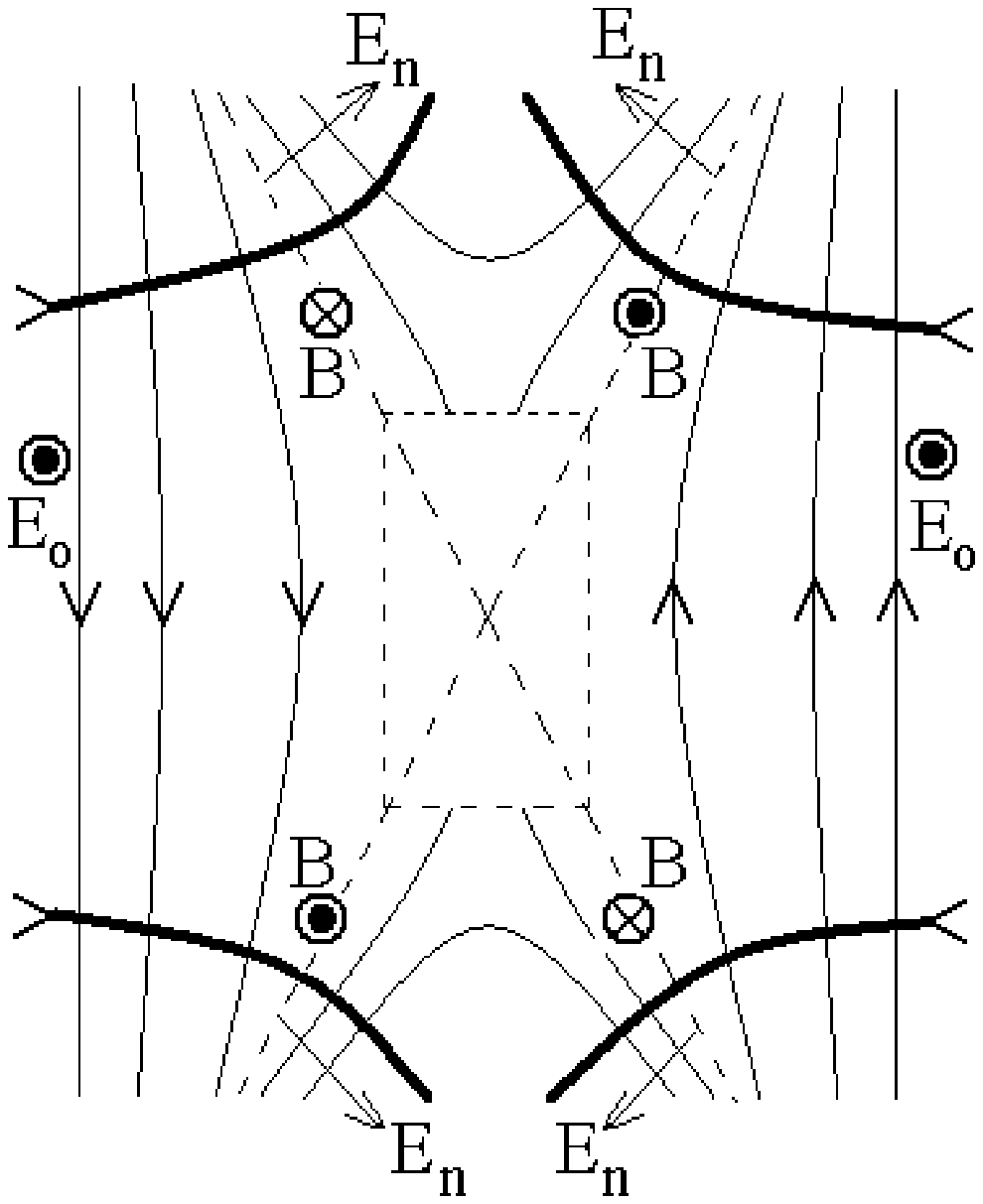}\\
\centering Figure 8
\label{fig08}
\end{figure}

     The discussion thus far has left many unanswered questions.  For example:

\begin{itemize}
\item \textbf{How do the ions and electrons move to create the current, $\mathbf{j}$?}
\item \textbf{How are the ions accelerated to the Alfven speed?}
\item \textbf{How is the parallel electric field in the central region generated?}
\end{itemize}

\noindent These questions will be discussed qualitatively through application of the Generalized Ohm's Law as derived from the two-fluid equations of motion, which are \citep{spitzer56}:\\

\noindent For ions
\begin{eqnarray}
n_{i} m_{i} [\partial \mathbf{U}_{i} / \partial t + (\mathbf{U}_{i} \cdot \nabla) \mathbf{U}_{i}] = \nonumber \\
n_{i} Z e (\mathbf{E} + \mathbf{U}_{i} \times \mathbf{B}) - \nabla \cdot \mathbf{P}_{i} + \mathbf{P}_{ie}
\label{eqn08}
\end{eqnarray}

\noindent For electrons
\begin{eqnarray}
n_{e} m_{e} [\partial \mathbf{U}_{e} / \partial t + (\mathbf{U}_{e} \cdot \nabla ) \mathbf{U}_{e}] = \nonumber \\
-n_{e} e (\mathbf{E} + \mathbf{U}_{e} \times \mathbf{B}) - \nabla \cdot \mathbf{P}_{e} + \mathbf{P}_{ei}                                                          \label{eqn09}
\end{eqnarray}

\noindent where
\begin{itemize}
\item $\mathbf{U}_{i}$ and $\mathbf{U}_{e}$ are the ion and electron velocities averaged over unit volumes
\item $n_{i}$ and $m_{i}$ are the ion density and mass, respectively
\item $n_{e}$ and $m_{e}$ are the electron density and mass, respectively
\item $P_{i}$ and $P_{e}$ are the ion and electron pressure tensors
\item $P_{ie}$ and $P_{ei}$ are the momentum transferred to ions (electrons) from electrons (ions) per unit volume per unit time.
\end{itemize}

\noindent Subtraction of (\ref{eqn09}) from (\ref{eqn08}) assuming:
\begin{itemize}
\item Neglect of quadratic terms
\item Electrical neutrality (which means that the electron and ion densities are equal everywhere except in Maxwell's equation for the divergence of the electric field)
\item Ignoring $m_{e}$/$m_{i}$ terms
\end{itemize}
gives the Generalized Ohm's Law

\begin{eqnarray}
\mathbf{E} + \mathbf{U}_{i} \times \mathbf{B} = \mathbf{j} \times \mathbf{B} / en -
\nabla \cdot \mathbf{P}_{e} / en + \nonumber \\
(m_{e} / ne^{2}) \partial \mathbf{j} / \partial t + \eta \mathbf{j}
\label{eqn10}
\end{eqnarray}

\noindent where $\eta$ is the Coulomb or anomalous resistivity.  Equivalently, because $\mathbf{j} / ne = \mathbf{U}_{i} -  \mathbf{U}_{e}$,   

\begin{equation}
\mathbf{E} + \mathbf{U}_{e} \times \mathbf{B} = -\nabla \cdot \mathbf{P}_{e} / en + (m_{e}/ne^{2})
\partial \mathbf{j} / \partial t + \eta \mathbf{j}
\label{eqn11}
\end{equation}

     For the current to be non-zero, the ions and electrons must move differently, i.e., the left side of equation \ref{eqn10} and/or \ref{eqn11} must be non-zero.  Because the $\mathbf{j} \times \mathbf{B}$ force of equation \ref{eqn10} acts on ions and because the ion inertial length, $c/\omega_{pi}$, and the ion gyroradius are much larger than the electron inertial length, $c/\omega_{pe}$, and the electron gyroradius, the ion bulk motion deviates from $\mathbf{E} \times \mathbf{B} / B^{2}$ on the larger spatial scale, called the ion diffusion region.  Meanwhile, the electron motion must include $\mathbf{E} \times \mathbf{B} / B^{2}$ flow perpendicular to the plane of Fig.\ 8 and into the figure because electrons provide the major part of the out-of-plane current.  This requires an in-plane electric field, $E_{n}$, perpendicular to the magnetic field and pointing towards the center of Fig.\ 8 from both sides in the ion diffusion region.

     Ions entering from the left or right move predominantly within the plane to be accelerated by $E_{n}$ and the $\mathbf{j} \times \mathbf{B}$ force to the opposite side of the central region where they decelerate, turn around, and again accelerate.  They bounce between the two sides to eventually exit vertically at the Alfven speed.  Because $en E_{n}  \approx jB$ according to equation \ref{eqn10}, and because the sum of the two forces times $c/\omega_{pi}$ is about the energy $B^{2}/\mu_{o}$ gained by a unit volume of fluid, the electric field is deduced as $E_{n} \approx v_{alfven} B/2$.  This is $\sim$20 mV/m at the sub-solar magnetopause, which is much larger than the fraction of a mV/m reconnection field, $E_{o}$, observed in space and in simulations.  $E_{n}$ and the ion counter-streaming and acceleration have been observed in space \citep{mozer02,wygant05}.  It is also noted that the ions are accelerated in a region having the ion scale and not in the much smaller electron diffusion region that is discussed below.
                     
               The electric field, $E_{n}$, existing along the dashed X in Fig.\ 8 over a scale size of $c/\omega_{pi}$, has a non-zero divergence so it must result from charge separation.  This requires that electrons move along the magnetic field lines in the plane of Fig.\ 8 and into the dashed box to create the charge separation.  The parallel (to $\mathbf{B}$) in-plane electric field in the dashed box accelerates these electrons to the outgoing ion bulk velocity, after which they are ejected into the upper and lower central regions.  This in-plane electron current causes the out-of-plane quadrupolar magnetic field that is illustrated in Fig.\ 8 and that has been observed in space \citep{mozer02} and in the lab \citep{ren05}.  The $\mathbf{j} \times \mathbf{B} /en$ term in equation \ref{eqn10}, the quadrupolar magnetic field, and the bipolar electric field constitute the essence of Hall MHD physics.  

     Fig.\ 9 shows a sub-solar magnetopause crossing by the Polar satellite in which the Hall MHD terms are quantitatively present.  Panel (d) gives the reconnection magnetic field which changes from $-80$ nT in the magnetosheath to $+80$ nT in the magnetosphere.  Panel (c) gives the out-of-plane magnetic field component which shows both a small guide field and the quadrupolar magnetic field.  Panel (e) gives the bipolar normal electric field.  It is noted that the tangential electric field of panel (f) is positive which means that magnetic field lines and plasma were flowing toward the center of the magnetopause from each side.  It is also noted that the bipolar electric field is an order-of-magnitude larger than the tangential electric field.

\begin{figure}
\centering \includegraphics[width=2.3in]{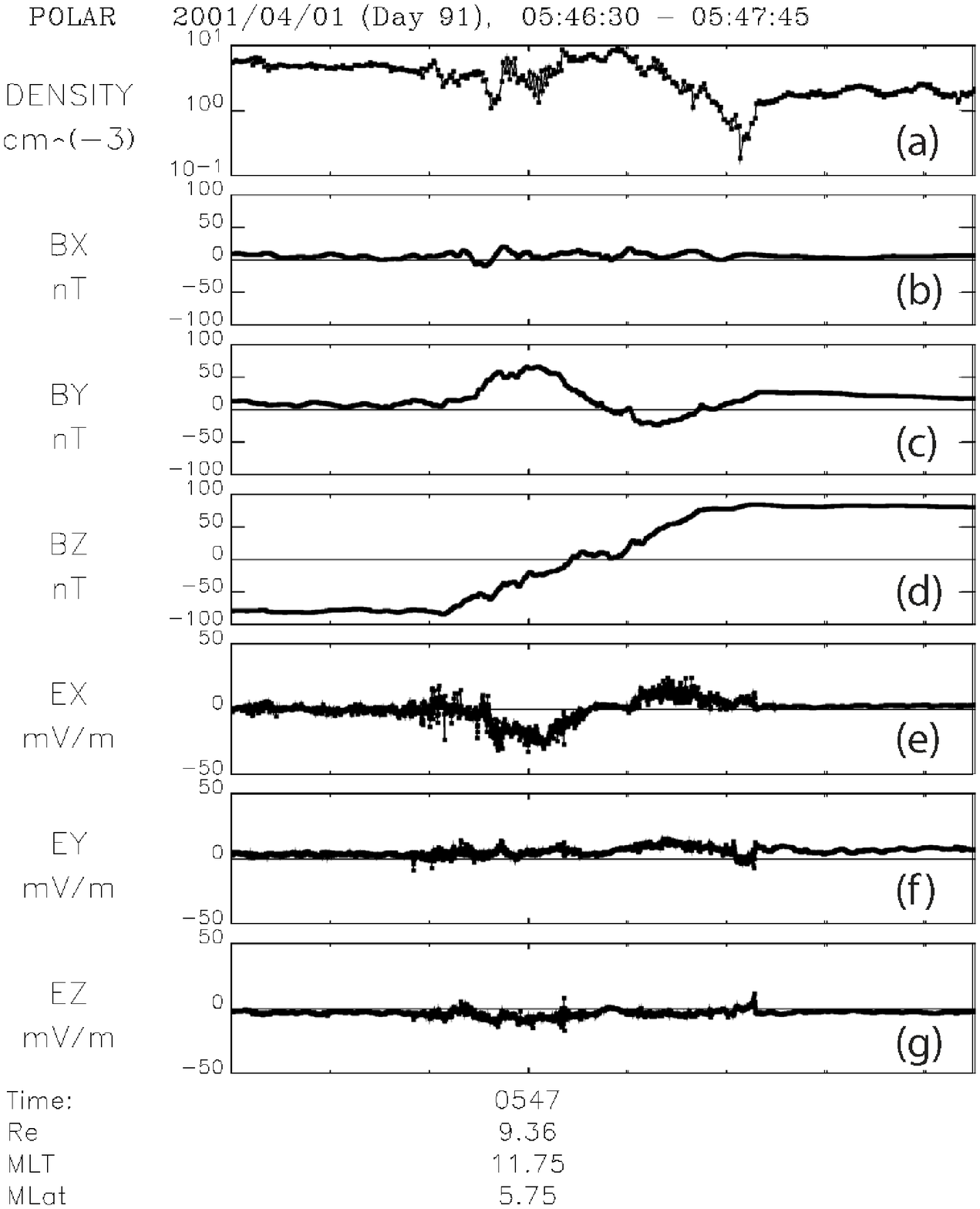}\\
\centering Figure 9
\label{fig09}
\end{figure}

     It is important to note that less than 1\% of satellite sub-solar magnetopause crossings exhibit clear signatures of the quadrupolar magnetic field and bipolar electric field of Fig.\ 8 \citep{mozer02}.  This is undoubtedly due to time variations, real geometries that differ from the idealized two-dimensional geometry of equal and opposite magnetic fields, spacecraft crossings at distances far from the center of the reconnection region, etc.\\

       \textbf{What is the nature of the parallel electric field inside the dashed box of Fig.\ 8?}  The right side of equation \ref{eqn11} must differ from zero in this region because of the parallel electric field on the left side.  Thus, the parallel electric field must be associated with either the time derivative of $\mathbf{j}$ (inertial effects), the divergence of the pressure tensor, or finite resistivity.  Experimental data suggests that it is associated with the pressure tensor \citep{mozer02,mozer05a}, but this is an important issue that remains to be fully resolved.  

     It is noted that the parallel electric field cannot be balanced by an isotropic pressure gradient.  This is because $\nabla \times \mathbf{E}_{\parallel}$ is non-zero and the isotropic pressure gradient that would balance $\mathbf{E}_{\parallel}$ has a curl that is zero (because the curl of the gradient of any scalar is zero).  Thus, off-diagonal terms in the pressure tensor are required for the pressure term to balance the parallel electric field.  Consideration of these off-diagonal terms requires a kinetic theory that is beyond the scope of the present discussion.  
 
     An example of an electron diffusion region observed in the sub-solar magnetopause is given in the four second plot of Fig.\ 10.  Panel (b) gives the reconnecting component of the magnetic field which changed in steps as the spacecraft passed from the magnetosheath to the magnetosphere.  The stepwise nature of this component shows that the current was filamentary.  In the largest current filament, the plasma density of panel (a) decreased by a factor of two while there was a large, electric field with a duration of about 0.075 seconds, as illustrated in the bottom three panels.  This structure had a spatial dimension of $\sim c/\omega_{pe}$.  Panel (e) shows that the parallel electric field in this electron diffusion region had a magnitude of $\sim$7 mV/m.  It is also noted that the magnetic field strength at the time of the event was $\sim$60 nT and the electron beta was much less than one.

\begin{figure}
\centering \includegraphics[width=2.3in]{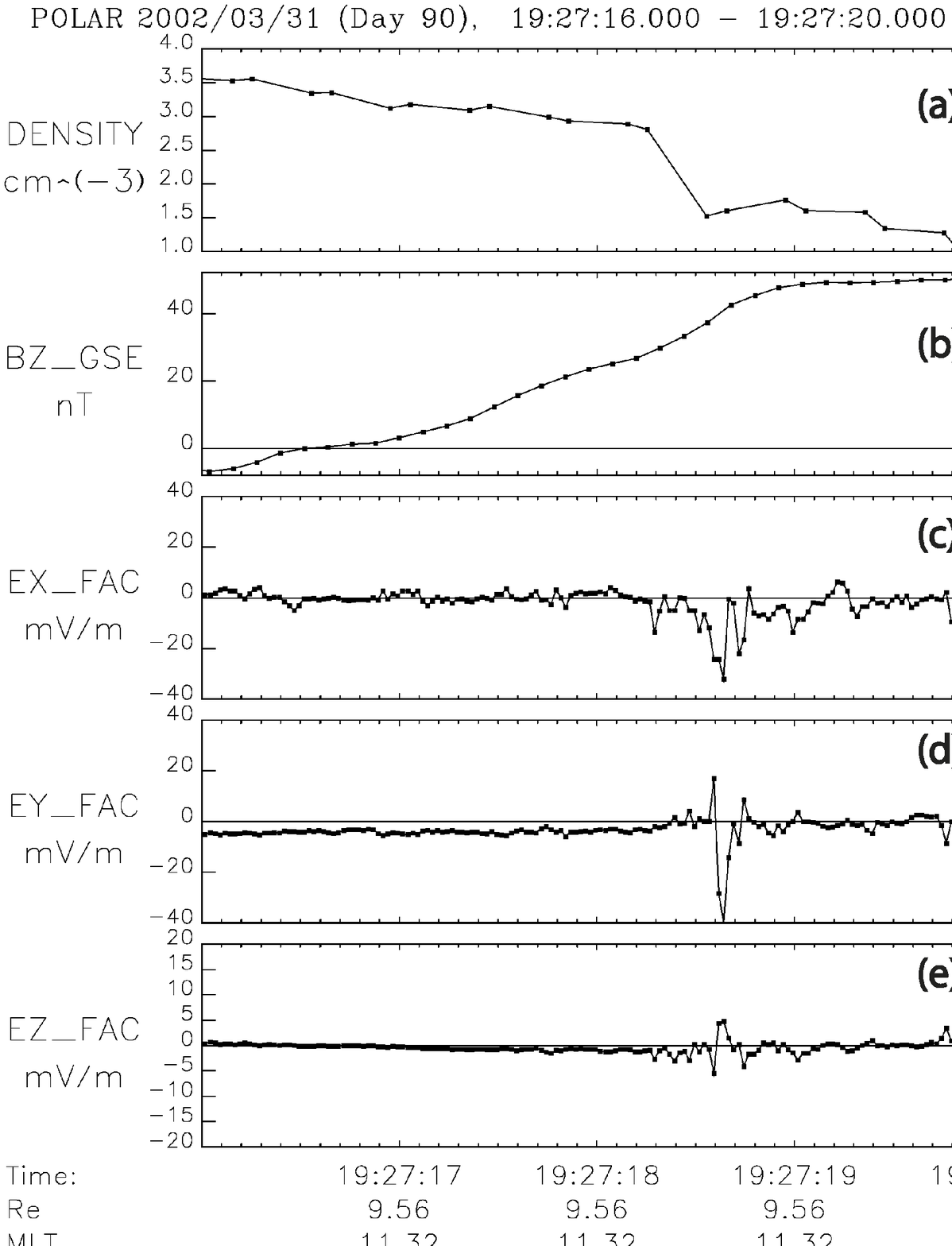}\\
\centering Figure 10
\label{fig10}
\end{figure}

     As suggested by the steady state nature of the fields in the ion diffusion region, it is plausible that the inertial terms of equation \ref{eqn11} are small in this region.  However, in the dashed box of Fig.\ 8, they are probably important because a non-zero $\nabla \times \mathbf{E}_{\parallel}$ requires non-zero terms in $\nabla \times \mathbf{E}$ and a non-zero $\partial \mathbf{B} / \partial t$, unless other terms in $\nabla \times \mathbf{E}$ exactly cancel the non-zero terms, which is unlikely.  

     The scale size for which the terms on the right side of equation \ref{eqn11} become important is $\sim c/\omega_{pe}$ \citep{vasyliunas75}.  A single region of this size is frequently drawn at the crossing of the dashed X in Fig.\ 8 to represent the electron diffusion region in which electrons are demagnetized.  From such a cartoon, it is sometimes incorrectly concluded that the magnetic field must be zero and the plasma beta must be large in the electron diffusion region.  Because physical conclusions should not be obtained from cartoons, it is important to consider that the dashed box may contain a large number of electron scale structures, each of which is located where the magnetic field is non-zero and each of which contains a parallel electric field.  This possibility is strongly supported by theoretical ideas \citep{tajimashibata02},  computer simulations \citep{onofri04,daughton06} and experimental data \citep{mozer03a,mozer05a,mozer05b}.  The locations of 117 electron diffusion regions found in the sub-solar magnetopause in searching three years of Polar data are shown in Fig.\ 11.  The sub-solar point is at the center of the plot which spans a local time of eight hours and a latitude of $\pm 40$ degrees.  This data suggests that electron diffusion regions occur frequently and at all locations within the magnetopause.

\begin{figure}
\centering \includegraphics[width=2.3in]{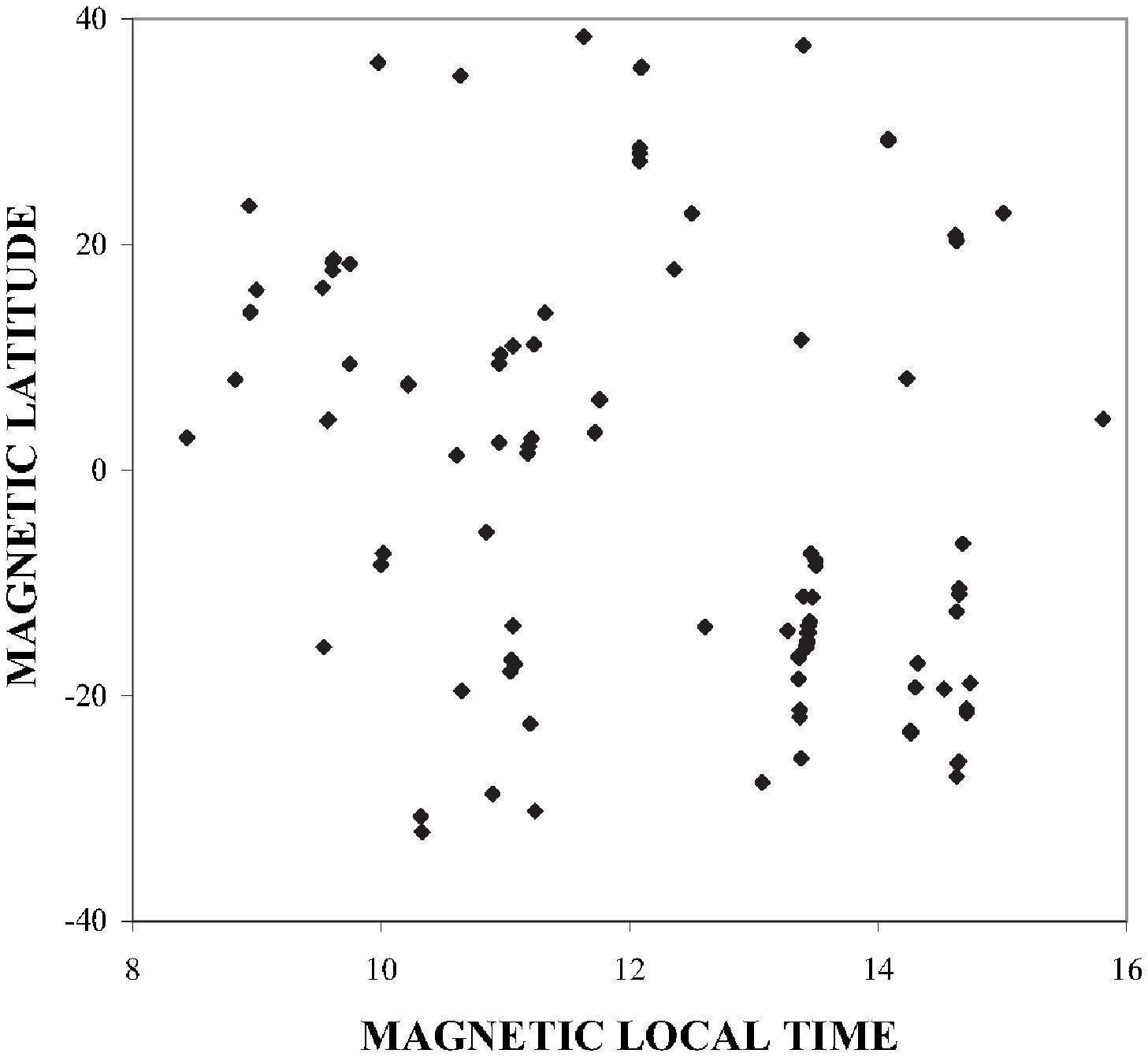}\\
\centering Figure 11
\label{fig11}
\end{figure}

     The reconnection rate is proportional to the number of magnetic field lines crossing into the reconnection region per unit area per second.  Because the density of field lines is $B$ and their velocity is $\mathbf{E}_{o} \times \mathbf{B} / B^{2}$, the rate of field line entry is the density times the velocity, or $E_{o}$.  This is why the tangential electric field is sometimes referred to as the reconnection rate.\\

     \textbf{What does sub-solar reconnection do for magnetospheric physics?}  Because most of the accelerated plasma escapes into the solar wind, it may appear that dayside reconnection is not important.  This conclusion is incorrect because reconnection also modifies the magnetic field connectivity to produce a long tail in the anti-sunward direction, as illustrated in Fig.\ 7.  In the southern hemisphere tail, the magnetic field points away from the earth and in the northern hemisphere tail, it points earthward.  In the presence of a dawn-to-dusk electric field, the Poynting vector along the tail surface is everywhere inward, so its integral over this closed surface provides an energy source for driving magnetospheric processes.  For a tail that is 300 earth radii long and 30 earth radii in diameter, the energy input associated with a magnetic field of 10 nT and a cross-tail electric field of 0.5 mV/m is $\sim 2 \times 10^{12}$ watts.  This is more than sufficient to drive all known magnetospheric processes and it is more than four orders-of-magnitude larger than the energy directly available from sub-solar reconnection.  The EMF that provides this energy is the kinetic energy of the solar wind, which slows as it moves along the tail.

     For the sub-solar region, it may seem that the reconnection rate need only be large enough to produce the long tail.  However, a large Poynting flux into the tail implies a large $\mathbf{E} \times \mathbf{B} / B^{2}$ flow of magnetic field lines into the tail and these field lines must also reconnect.  Because the average reconnection on the dayside and in the tail must be equal, a large Poynting flux into the tail must be accompanied by a large sub-solar reconnection rate.

      In summary, topics for future research on magnetic field reconnection include:

\begin{itemize}
\item Microphysics in the electron diffusion region.  Understanding the relative importance of inertial effects, resistivity, and the gradient of the pressure tensor.
\item Processes that result from the reconnection geometry and that accelerate ions and electrons to high energies on short time scales.
\item Wave modes associated with reconnection and the subsequent particle acceleration.
\item The mechanism for generating parallel electric fields in the electron diffusion region.
\item The importance of electron diffusion regions that are observed throughout the magnetopause as well as in the magnetosphere, the magnetosheath, and the bow shock \citep{mozer06}.
\end{itemize}

\noindent \textbf{Acknowledgements}

     The author thanks C.-G. Falthammar for many helpful discussions over the years, and his Berkeley colleagues for many helpful comments.  This work was supported by NASA Grants NNG05GC72G and NNG05GL27G.

\end{document}